# TinySearch- Semantics based Search Engine using Bert Embeddings

Manish Patel
*Computer Science and Engineering*
*Texas A&M University*
College Station,US
manishp@tamu.edu

*Abstract*—Existing search engines use keyword matching or tf-idf based matching to map the query to the web-documents and rank them. They also consider other factors such as page rank, hubs-and-authority scores, knowledge graphs to make the results more meaningful. However, the existing search engines fail to capture the meaning of query when it becomes large and complex. BERT, introduced by Google in 2018, provides embeddings for words as well as sentences. In this project, I have developed a semantics-oriented search engine using neural networks and BERT embeddings that can search for query and rank the documents in the order of the most meaningful to least-meaningful. The results shows improvement over one existing search engine for complex queries for given set of documents.

*Index Terms*—Deep Neural Networks(DNN), Bidirectional Encoder Representations from Transformer (BERT), cosine-similarity, Long-Short Term Memory (LSTM), Siamese LSTM.

## I. INTRODUCTION

According to this survey [10], there are over 1.5 billion websites in the world, out of which around 250 million are active i.e. are still serving unique contents. Thus, searching for content on web is akin to finding needle in the haystack. But the good news is that we have search engines to help us fulfill our information need. Google, Yahoo, Baidu, Bing, DuckDuckGo, Yandex are a few popular examples. These search engines use a plethora of strategies to rank the documents some of which are- keyword matching of query with documents [1], tf-idf based vector space model and BM25 based vector-space model. To increase the precision i.e. to increase the relevance scores of pages which are shown at top, concepts such as PageRank, HITS, Knowledge Graph are used by the search engines. Moreover, with the increase in number of mobile phones and other GPS devices, search engines leverages location, time and other features to further refine the results.

However, search engines do not perform well if the query becomes too big and complex. For example for this query- "Find all the faculty members in deep learning field that are also experts in information theory", I find that three search engines- Google, Yahoo and Bing, performed the keyword based searching as seen in figure 1,2,3. The snippets highlight keywords such as 'faculty members','deep learning','field','information theory', 'learning', 'faculty'. Out of these three search engines result, I find Google's result more relevant than other Yahoo and Bing. Google's first two results provide links to web pages of professors who are experts in both fields, third result points to web page that mentions a 'Member who was working in quantum information theory before he joined Facebook AI Research', fourth link points to Deep Learning field in general and fifth link points to 25 Machine Learning books some of which covers ' concepts such as probability theory and information theory and describes deep learning techniques used in industry'.

Yahoo's first search results points to a web page of John Hopkins University that describes 'Spring 2019 courses taught by faculty members'. The courses include 'Deep Learning' and 'Computation Biology and Bioinformatics using information theory and markov modelling' which are most relevant to the query at hand. The fourth results points to a page containing interview of a new faculty member in Biomedical data science department who teaches a course called 'Deep Learning in Genomics and Biomedicine' and he has few fellow professors in his institute who teach 'information theory'. The second, third and fifth results of Yahoo are nowhere related to the given query.

Bing, on the other hand, provides link to the John Hopkins page as first result, interview of new faculty member as second result-both of these are discussed in the above paragraph. The third and fifth results are about faculty members in general and the fourth result is about 'Deep Learning and Neural Networks' course on 'Coursera taught by a faculty member from NYU'.

We can see from the above results that search engines do not perform well on complex queries such as the one above. They fail to notice that 'deep learning' and 'neural networks', 'faculty members' and 'professors' are synonymous terms. In general, as the length of query grows, the precision of results returned from search engines drop.

## II. PRELIMINARIES

*A. Neural Networks*

A Deep Neural Networks (DNN) are part of a family of machine learning methods based on learning data representations. A DNN can be denoted as a function F: X → Y, where X represents the input space and Y the output space. It consists of a number of connected layers, and links between layers are weighted by a set of weight-matrices. The training phase of a DNN is to identify the numerical values of the weight-matrices. The training procedure utilizes a large dataset of known input-output pairs and defines a loss function representing the differences between the predictions and the true labels. The training phase seeks to minimize the loss function by updating the parameters using the backpropagation technique.

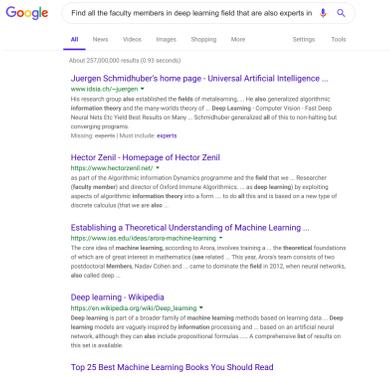

Fig. 1. Google Search results for the query

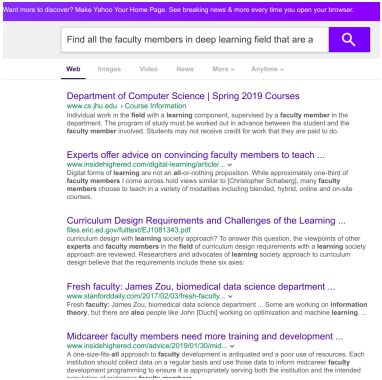

Fig. 2. Yahoo Search results for the query

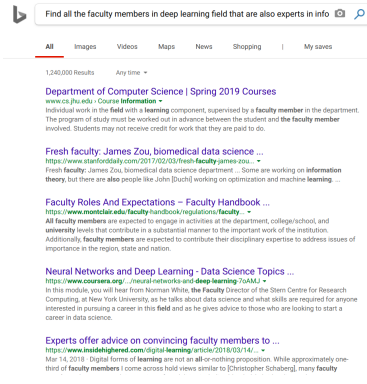

Fig. 3. Bing Search results for the query

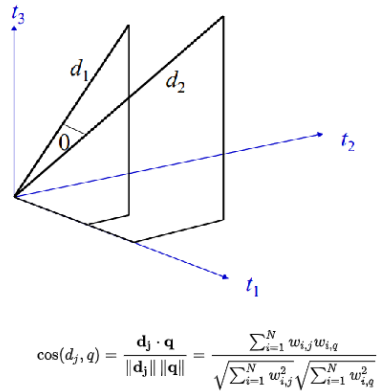

Fig. 4. Cosine similarity in vector space model

*B. BERT*

BERT or Bidirectional Encoder Representations from Transformers [4] is a language representation model developed by Google in 2018 to pre-train deep bidirectional representations by jointly conditioning on both left and right context in all layers. This is done by using two novel unsupervised prediction tasks described as under.

**Masked LM:** Masked Language Model allows us to train a deep bidirectional representation by taking a straightforward approach of masking some percentage of the input tokens at random, and then predicting only those masked tokens. This is shown in figure II-B.

**Next Sentence Prediction** Many important downstream tasks such as Question Answering (QA) and Natural Language Inference (NLI) are based on understanding the relationship between two text sentences, which is not directly captured by language modeling. In order to train a model that understands sentence relationships, the authors of the paper pre-train a binarized next sentence prediction task that can be trivially generated from any monolingual corpus as shown in the figure II-B.

The advantage of BERT is that it can be fine-tuned for downstream tasks such as Sentence Pair Classification, Sentence Tagging, Question Answering etc, to provide embedding or encoding that contain semantics of the original sentence.

*C. Vector Space Models*

Vector space model is an algebraic model for representing documents and query as vector of identifiers. Each term in the document represents a dimension,so that for d terms a dimensional space is created. Then document and query are vectors in this d-dimensional space as shown in figure 4. The relevance ranking of documents is done by computing the similarity score which is the angle between the vectors in this d-dimensional space.This is done by taking the dot product of two vectors and normalizing it to make it independent of the length of documents.

## III. RELATED WORK

Kassim et al. [3] proposed the Semantic Search Engine which consists of Ontology development, Ontology Crawler, Ontology Annotator , Web Crawler, Semantic Search and Query Processor. My work is different from this work since I

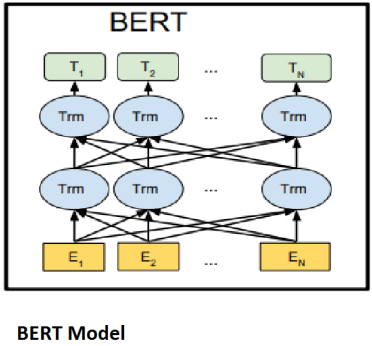

Fig. 5. BERT model from Google

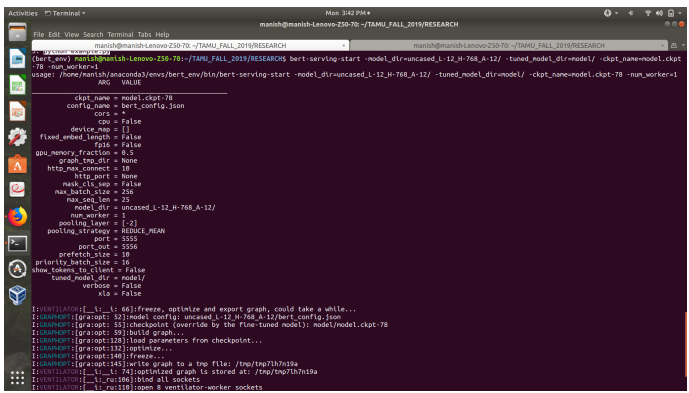

Fig. 6. BERT Server

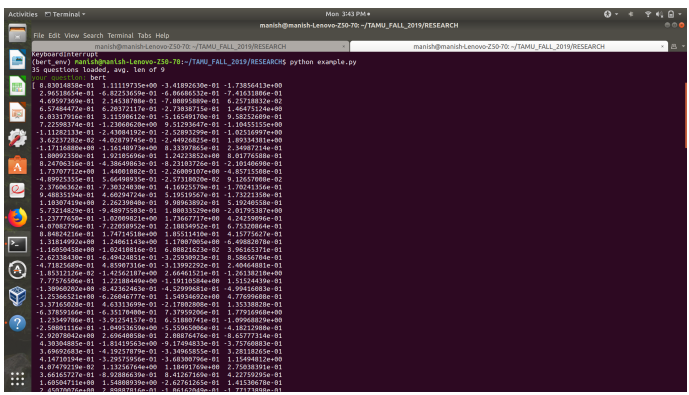

Fig. 7. Embedding from BERT server

am not using Ontology to store structure of words and create domain related information structures.

Webler [4], in this thesis work art2vec, has proposed a semantic search engine for tagged artworks based on word embeddings. My work is similar to it but is different since I have used BERT embeddings rather than word2vec or Doc2vec embeddings and also I have used a neural network to further find the similarity between two document vectors.

Josh Taylor [5] has developed a semantic search engine using ELMo embeddings while Han Xiao [6] has developed the search engine using BERT embeddings. Both of the works used cosine similarity to compute the similarity score of query and documents. My work uses BERT embeddings but uses neural network to find the similarity score.

IV. DESIGN

The architecture of TinySearch is shown in the figure 8. It consists of a) BERT server which performs the encoding of the text and returns an embedding vector, b) a neural network model that return similarity score of two embedding vector, and c) a GUI which takes in an input query from the user, get its embedding from the BERT server, computes its similarity to the documents and returns the most relevant 5 documents. The detailed design and implementation of each components is discussed below.

*A. BERT Server*

The BERT server [6] is an open source highly scalable sentence encoding service based on Google BERT from Han Xiao. It allows one to map a variable-length sentence to a fixed-length vector. In its raw form, it uses pre-trained uncased BERT model of 12 layers and produces an embedding vector of length 768. However, I have fine-tuned the BERT model on MRPC dataset [9] and used the [CLS] token to get the embedding of the entire document. This is unlike taking the average of word-vector embeddings where the most frequent word or most dominant idea may dominate the entire embedding. According to the original paper [7], the first token ([CLS]) is a special embedding. The final hidden state corresponding to this token is used as the aggregate sequence representation for classification tasks. Figure 6 shows the BERT server at working and 7 shows the output embedding vector of length 768.

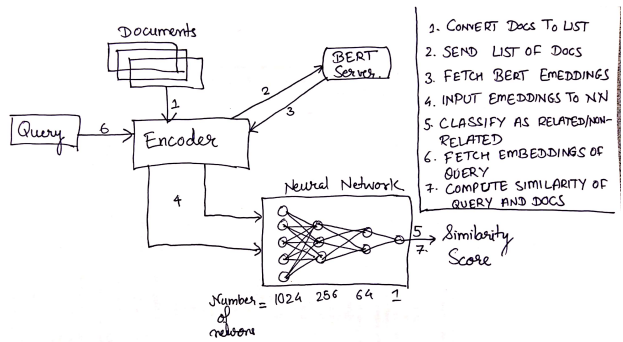

Fig. 8. TinySearch Architecture

### B. Neural Network Model

I have implemented a deep neural network that takes two embedding vectors, concatenates them and tries to find the similarity score of the sentences. To train this model, I have used the quora_question pairs dataset [8]. The dataset consists of almost 400000 pairs but I have used only 100000 for training. The first step in training is to generate the embeddings of question pairs. For this I send the list of question pairs to BERT server, which encodes it and returns a list of the embeddings. I save this embeddings using pickle to avoid repetition of this task in future. Then I load the embeddings and pass it to the two Input Layers of shape (768,1). Then I concatenate these inputs and connect it with a dense layer of size 1024. Then I use two more Dense layers, one of size 256 and another of size 64. The reason behind increasing and then decreasing it that model may first try to learn more specific features and then general features.However, to reduce the overfitting, I have used a dropout rate of 0.5 Finally, I have used sigmoid activation to generate value between 0 and 1.

```
input1 = Input(shape=(768,1))
input2 = Input(shape=(768,1))

x = keras.layers.concatenate([input1,input2],
   axis=-1)
x = Dense(1024,activation='relu') (x)
x = Dropout(0.5) (x)
x = Dense(256,activation='relu') (x)
x = Dropout(0.5) (x)
x = Dense(64,activation='relu') (x)
output = Dense(1,activation='sigmoid') (x)

model =
   Model(inputs=[input1,input2],outputs=output)
model.summary()
model.compile(optimizer='rmsprop',
   loss='binary_crossentropy',
   metrics=['acc'])
```

```
history=model.fit([train_vec1, train_vec2],
   train_label,
   epochs=30,batch_size=200,
   validation_split=0.3)
```

### C. Graphical User Interface

To implement the GUI 11, I have used Tkinter library of Python. The GUI displays 14 documents which are representative of web pages for a search engine. I have used Google search engine to collect the documents that are displayed as result of the queries. The intuition behind it is that the google snippets displayed for a web page explains why it was picked from all the web pages. Then I have displayed a 'search text' button where user can enter its query. When user hits find button the query text is first sent to the BERT server to generate its embedding vector. Then this embedding vector along with embedding vectors of documents are sent to the neural network model which presents a score between 0 and 1. The documents are then sorted in descending order and top 5 results are displayed to the user.

## V. EVALUATION

Evaluation of three components of TinySearch has been done and is discussed as under.

### A. BERT Server

Fine tuning the BERT-server on Microsoft Research Paraphrase corpus that consists of 3600 examples lead to an accuracy of around 84%. Running it multiple times resulted in 3 to 4 percent variations which is acceptable since dataset is small.

```
***** Eval results *****
 eval_accuracy = 0.845588
 eval_loss = 0.505248
 global_step = 343
 loss = 0.505248
```

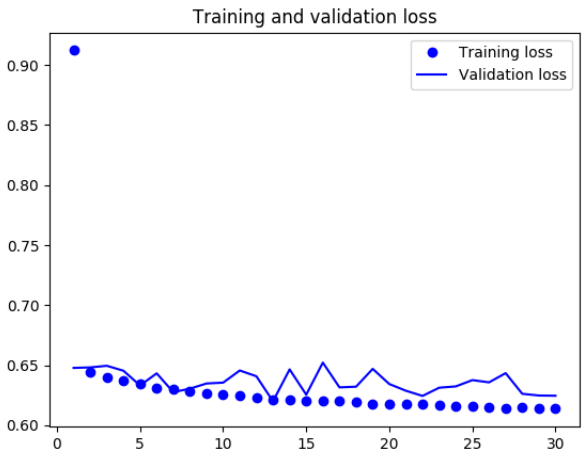

Fig. 9. Validation Loss

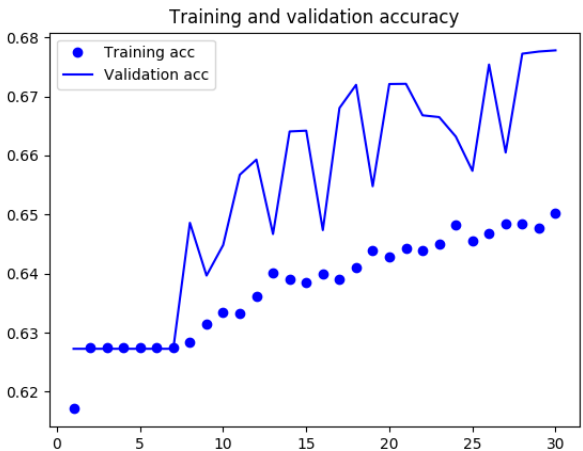

Fig. 10. Validation accuracy

### B. Neural Network Model

I have trained the model for 30 epochs with a batch size of 200. The validation accuracy kept on fluctuating but on average it was around 67%. The validation loss became almost constant after reaching 63%. Figure 9 and 10 shows the validation loss and accuracy.

### C. Graphical User Interface

To evaluate the results displayed for a query is a challenging task since there is no baseline to compare the model. However, precision and F1 score are well suited for ranking purposes and I have used these two metrics to evaluate the results displayed. Note that I have assigned "gold score" which denotes the actual relevance of the documents to the given query, myself based on their actual meaning without being influenced by scores of search engines or results of GUI itself.

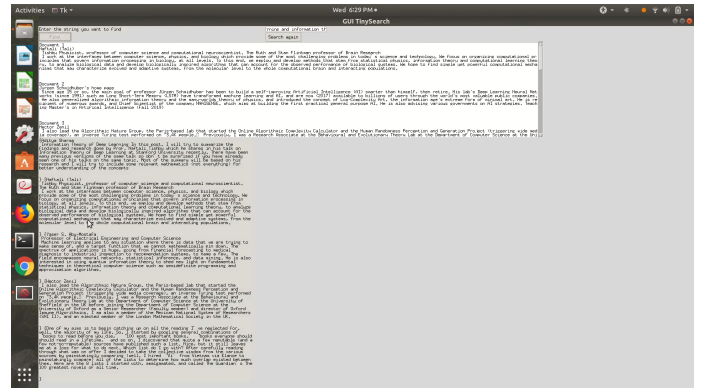

Fig. 11. GUI

The three types of queries made to the GUI and their evaluation on the metric are discussed as under:

```
Query1: Find all the faculty members who
work in deep learning and information theory

True Positives= 4
False Positives= 1
True Negatives= 8
False Negatives= 1

Precision = 0.8
Recall=  0.8
F1 score = 8

Query2: football in usa

True Positives= 2
False Positives= 3
True Negatives= 8
False Negatives= 3

Precision = 0.4
Recall=  0.4
F1 score = 4

Query1: must read english classic
books of all time

True Positives= 2
False Positives= 3
True Negatives= 6
False Negatives= 3
```

```
Precision = 0.4
Recall=   0.4
F1 score = 4
```

## VI. Conclusion

Finding semantic similarity of two sentences has always been a challenge in NLP. BERT, introduced by Google in 2018, marks the beginning of new era in NLP because of its ability to perform various NLP tasks better than previous techniques. In this project, I found that using queries of larger length results in better Precision and Recall than shorter queries. This makes sense because shorter queries can be ambiguous in meaning and also can't point to a particular document in the vector space since its embedding vector will be far from embedding vector of all the documents. Please find the demo link in references.

## VII. Future Work

In future, I would like to extend this project by using Siamese LSTM neural network with Manhattan distance. Also, I would use BM25 as a baseline model. Furthermore, I can use ensemble method of BM25 for shorter queries and BERT for longer queries to make the search engine more accurate.

## Acknowledgment

This work was done as part of the project for the CSCE 636 (Spring 2019) course on Neural Networks taught by Dr. Andrew Jiang. I thank Dr. Jiang for his feedback on my project.